\documentclass[aps,pra,twocolumn,showpacs,floatfix,superscriptaddress]{revtex4-1}
\usepackage{graphicx}
\usepackage{hyperref}
\usepackage{amsmath}
\usepackage{amssymb} 
\usepackage{color}

\newcommand{\mean}[1]{\langle{#1}\rangle} 

\newcommand{\ud}[1]{{#1^{\dagger}}}

\begin{document}
\flushbottom
\title{Optimization of photon correlations by frequency filtering}

\author{Alejandro Gonz\'alez-Tudela}
\affiliation{Max--Planck Institut f\"ur Quantenoptik, 85748 Garching, Germany}

\author{Elena del Valle}
\affiliation{Departamento de F\'isica Te\'orica de la Materia Condensada, \\ Universidad Aut\'onoma de Madrid, 28049 Madrid, Spain}

\author{Fabrice~P.~Laussy}
\affiliation{Departamento de F\'isica Te\'orica de la Materia Condensada, \\ Universidad Aut\'onoma de Madrid, 28049 Madrid, Spain}
\affiliation{Russian Quantum Center, Novaya 100, 143025 Skolkovo, Moscow Region, Russia}

\begin{abstract}
  Photon correlations are a cornerstone of Quantum Optics. Recent
  works~[NJP \textbf{15} 025019~\&~033036 (2013), PRA \textbf{90}
  052111 (2014)] have shown that by keeping track of the frequency of
  the photons, rich landscapes of correlations are revealed. Stronger
  correlations are usually found where the system emission is weak.
  Here, we characterize both the strength and signal of such
  correlations, through the introduction of the ``frequency resolved
  Mandel parameter''. We study a plethora of nonlinear quantum
  systems, showing how one can substantially optimize correlations by
  combining parameters such as pumping, filtering windows and time
  delay.
\end{abstract}
\pacs{42.50.Ar,42.50.Lc,05.10.Gg,42.25.Kb,42.79.Pw}
\date{\today} \maketitle

\section{Introduction}

The quantum theory of optical coherence developed by Glauber in the
60s~\cite{glauber63a,glauber63b} revolutionized the field of Quantum
Optics by identifying photon correlations as the fundamental
characterization of light, instead of
frequency~\cite{glauber06a}. This is a great insight since coherence
had been understood for centuries as a feature of monochromaticity
while it is now understood in terms of factorizing correlators.  This
has been confirmed experimentally with the advance of new light
sources (such as the laser or single-photon sources~\cite{lounis05a})
as well as progress in photo-detection. In the quantum picture, the
frequency $\nu$ of light is linked to the energy~$E$ of its
constituting particles through Planck constant: $E=h\nu$.  The
standard approach of photon correlations has consisted so far
essentially in detecting photons from a light source as a function of
time, disregarding their frequency. Experimentally this is achieved
either with the original Hanbury Brown--Twiss~\cite{hanburybrown56a}
configuration or by detecting individual photons with a streak
camera~\cite{wiersig09a}. For stationary signals, the most important
photon correlation is measured by the second order correlation
function:
\begin{equation}
 \label{eq:eqg2}
 g^{(2)}(\tau)=\lim_{t\rightarrow \infty} \frac{\mean{\ud{a}(t)\ud{a}(t+\tau)a(t+\tau)a(t)}}{\langle(\ud{a}a)(t)\rangle\langle(\ud{a}a)(t+\tau)\rangle}
\end{equation}
with~$a(t)$ the light field annihilation operator of the system under
study at time~$t$. If the corresponding spectral shape is
singled-peak, the question of frequency correlations of the emitted
photons may appear a moot point. We will see shortly that it is
not. In many cases, nevertheless, the emission is multi-peaked and it
is then clear that Eq.~(\ref{eq:eqg2}), which correlates photons
regardless of which peak they originate from, is leaving some
information aside. It is natural to inquire what are the correlations
of each peak in isolation, or what are the cross-correlations between
peaks~\cite{cohentannoudji79a, dalibard83a}. Experimentally, this is
readily achieved by inserted filters in the arms of the Hanbury
Brown-Twiss configurations~\cite{akopian06a, hennessy07a, kaniber08a,
  sallen10a,ulhaq12a} or using a monochromator in a streak camera
set-up~\cite{arXiv_silva14a}. Theoretically, the Glauber correlator
must be upgraded to the so-called time and frequency resolved photon
correlations, \cite{dalibard83a,knoll86a, nienhuis93a}:
\begin{align}
 \label{eqg2omega2}
  & g_{\Gamma_1,\Gamma_2}^{(2)}(\omega_1,\omega_2;\tau)=\nonumber \\ & \lim_{t \rightarrow \infty}\frac{\mean{\ud{A}_{\omega_1,\Gamma_1}(t)\ud{A}_{\omega_2,\Gamma_2}(t+\tau)A_{\omega_2,\Gamma_2}(t+\tau)A_{\omega_1,\Gamma_1}(t)}}{\langle(\ud{A}_{\omega_1,\Gamma_1} A_{\omega_1,\Gamma_1})(t)\rangle\langle(\ud{A}_{\omega_2,\Gamma_2}A_{\omega_2,\Gamma_2})(t+\tau)\rangle}\,,
\end{align}
where $A_{\omega_i,\Gamma_i}(t)=\int_{-\infty}^t dt_1
e^{(i\omega_i-\Gamma_i/2) (t-t_1)} a(t_1)$ is the field detected at
frequency~$\omega_i$, within a frequency window~$\Gamma_i$, at time
$t$. We have recently developed a theory to compute such
correlations~\cite{delvalle12a} and introduced the concept of
``two-photon correlation spectrum'' (2PS) which, beyond correlating
merely peaks, spans over all the possible combinations of photon
frequencies~\cite{delvalle13a,gonzaleztudela13a}.  Landscapes of
correlations of unsuspected complexity are revealed as a result, which
are averaged out in standard photon detection or remain hidden when
constraining to particular (fixed) sets of frequencies.  The 2PS
enlarges the set of tools in multidimensional
spectroscopy~\cite{nardin13a,ra13a,nardin14a,gessner14a,schlawin14a}
and reveals a new class of correlated emission, that can be useful for
quantum information processing~\cite{sanchezmunoz14b,flayac14a},
enhance squeezing~\cite{grunwald15a} or for the study of the
foundations of quantum mechanics~\cite{arXiv_folman13a,arXiv_silva14a}.
When looking at the full picture, strong correlations turn out to
originate from photons not part of the spectral peaks, since a peak
results from a single-photon transition between two real
states. Various such photons have weak correlations and even when they
do, they are of a classical character. In contrast, collective
transitions that require two photons to undertake the emission are
strongly and non-classically correlated. Since they involve a virtual
state whose energy is not fixed, unlike for real states, they are
emitted at other frequencies than those of the
peaks~\cite{delvalle12a,delvalle13a,gonzaleztudela13a,sanchezmunoz14b}.

Recently, the 2PS of a nontrivial quantum emitter has been
experimentally observed~\cite{arXiv_peiris15a}, with spectacular
agreement with the theory and positively identifying in a rich
landscape of correlations the ``leapfrog emission'', i.e., between two
real states separated by an intermediate virtual one, as well as their
violation of the Cauchy Schwarz inequalities. The emitter was a
semiconductor quantum dot and the physical picture that of resonance
fluorescence in the Mollow triplet regime~\cite{mollow69a}.  Shortly
before that, a 2PS of spontaneous emission was measured from a
polariton condensate~\cite{arXiv_silva14a}, which features, however,
no quantum correlated emission and presents instead a simpler and
smooth landscape alternating bunching and antibunching as frequencies
get similar or far apart, due to the fundamental Boson
indistinguishability. These experiments confirm that the theory is
sound and robust and that the physics of photon correlation is ripe to
take advantage of the effects uncovered by their tagging with a
frequency. For instance, the mere Purcell enhancement of leapfrog
processes results in $N$-photon emitters~\cite{sanchezmunoz14a}.

A central theme of frequency engineering is the interplay between
signal and correlations. Correlated emission transiting by virtual
states is a high-order process and is therefore much less frequent
than direct emission. This brings the concern of the practical
measurement of a 2PS, since this requires measuring coincidences from
spectral windows where the system already emits very
little. Mathematically, this difficulty is concealed for both classes
of correlations, Eqs.~(\ref{eq:eqg2}) and (\ref{eqg2omega2}) alike, by
the normalization (denominator) which balances the intensity of the
coincidence emission (numerator), turning two vanishing numbers into a
finite ratio. In this text, we address this problem and revisit the
2PS to take into account the available amount of signal. To do so, in
Section~\ref{sec:freqMandel}, we introduce the frequency-resolved
Mandel parameter $Q_{\Gamma_1,\Gamma_2}(\omega_1,\omega_2;\tau)$ that
combines both correlations and emission intensity. In the light of
this new parameter, we revisit the two-photon correlation map at
$\tau=0$~\cite{gonzaleztudela13a} for several paradigmatic examples in
nonlinear quantum optics built around a two-level system. Namely, we
consider both its incoherent and coherent driving, the latter bringing
the system into the Mollow triplet regime, already graced with its
experimental observation~\cite{arXiv_peiris15a}. We also consider its
coupling to a cavity to realize the Jaynes-Cummings (JC) physics, as
well as the biexciton configuration found in, typically, quantum dot
systems. These systems are briefly introduced all along the paper, but
mainly to settle notations and we refer to the literature for the
concepts attached to them as well as for their relevance to our
problem.

Even the Mandel parameter does not fully capture the problematic of
the signal, since some correlations are so strong that they dominate
over the scarcity of emission. In
Section~\ref{sec:vieago15194608CEST2014}, we complement the
information of the available signal with an estimate of the measuring
time this supposes, defining a notion of valleys of accessible
correlations. There is considerable freedom added by filtering photons
when studying their correlations, and we explore various ways to
optimize them. Various approaches are illustrated for various systems,
focusing on the JC model in Section~\ref{sec:JC} and the biexciton
cascade in Section~\ref{sec:biexciton}. In the JC case, we study the
optimization with the intrinsic system parameters, namely the
cavity-photon lifetime and the pumping rate, while in the biexciton
case, we study the dependence on extrinsic parameters, namely, the
filters linewidth and/or time delay. Clearly, a comprehensive analysis
could be given along such lines to any system of interest. The present
text aims at illustrating such points in particular cases and leaves
it to future works to combine them in the cases where they will be
needed.

\section{Frequency-resolved Mandel parameter \label{sec:freqMandel}}

Mandel introduced for standard photon-correlations (that is, without
the frequency information) a parameter~\cite{mandel65a}, now bearing
his name, intended to correct for the previously mentioned
normalization issue: the balancing of two vanishing quantities that
provide a finite-value correlation which is, for all practical
purposes, not measurable since these quantities are those accessible
to the experiment. The ``Mandel parameter'' is defined, for a
stationary signal, as:
\begin{equation} 
  \label{Eqmandel}
  Q(\tau)=n_a (g^{(2)}(\tau)-1)\,,
\end{equation}
where $n_a=\lim_{t \rightarrow \infty} \mean{\ud{a}(t)a(t)}$ is the
steady state population of the detected mode. The offset by unity
makes the Mandel parameter negative when the light is quantum (in the
sense that it is sub-Poissonian and as such has no classical
counterpart). The product by~$n_a$ makes the normalization of
coincidences to the average signal instead of, as previously, to
uncorrelated coincidences. It conveys, therefore, a meaningful
information on the magnitude of the available signal.  Note that
$Q(0)=0$ results either from the lack of correlated emission
($g^{(2)}(0)=1$) or from too little emission ($n_a\rightarrow 0$). In
this way, the Mandel parameter really characterizes the amount of
correlated emission.

Following the spirit of Mandel, we introduce a frequency resolved
version:
\begin{multline} 
  \label{Eqmandelfreq}
  Q_{\Gamma_1,\Gamma_2}(\omega_1,\omega_2;\tau)=\\
  \sqrt{S_{\Gamma_1}^{(1)}(\omega_1)S_{\Gamma_2}^{(1)}(\omega_2)}(g_{\Gamma_1,\Gamma_2}^{(2)}(\omega_1,\omega_2;\tau)-1)\,,
\end{multline}
where $S_{\Gamma_i}^{(1)}(\omega_i)=\lim_{t \rightarrow \infty}
\langle(\ud{A}_{\omega_i,\Gamma_i} A_{\omega_i,\Gamma_i})(t)\rangle$
is now the steady state spectrum, which represents, physically, the
amount of photons passing through the filter of linewidth $\Gamma_i$
centered at $\omega_i$. Here it must be emphasized that while
$Q(\tau)<0$ is a sufficient condition to establish the quantum
character of the emission, as it corresponds to a Cauchy-Schwarz
inequality (CSI) violation, there is not such a straightforward
interpretation for the frequency resolved version that would read:
\begin{equation}
  \label{eq:vieago15204543CEST2014}
  [g^{(2)}_\Gamma(\omega_1,\omega_2,0)]^2<
  g^{(2)}_\Gamma(\omega_1,\omega_1,\tau)g^{(2)}_\Gamma(\omega_2,\omega_2,\tau)\,.
\end{equation}
Such violation of classical inequalities gives rise to their own
landscape of correlations~\cite{sanchezmunoz14b}. In contrast,
the anticorrelation in frequency, which we will qualify as ``frequency
antibunching'' in agreement with the literature~\cite{deutsch12a},
only reflects anti-correlations of intensities, which can be, or not,
linked to a quantum character of the emission.

\begin{figure}[!tb]
  \includegraphics[width=0.8\linewidth]{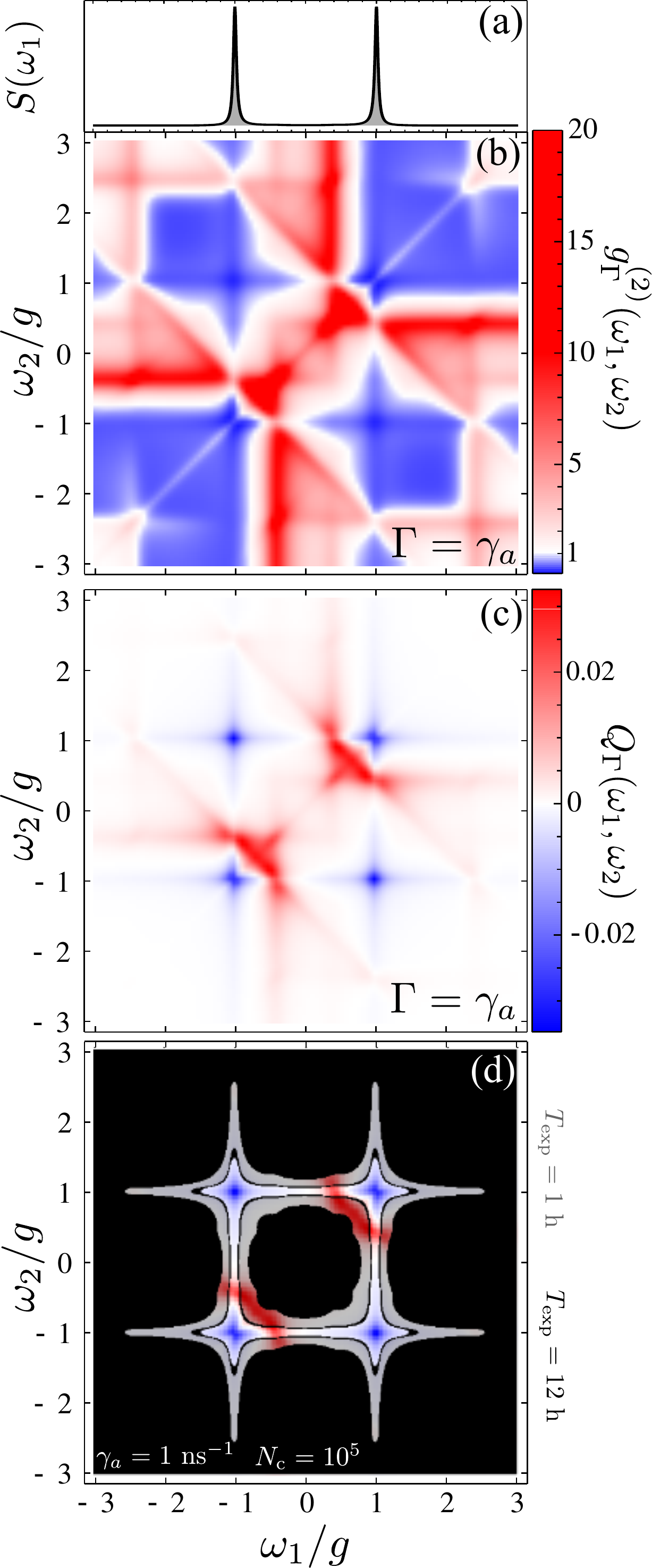}
  \caption{(Colour online) (a) One-photon spectrum for the JC system
    at low pumping, exhibiting the Rabi doublet of strong
    coupling. (b) The corresponding 2PS obtained by spanning
    $g_{\gamma_a}^{(2)}(\omega_1,\omega_2)$ over all frequencies (c)
    The corresponding frequency-resolved Mandel parameter
    $Q_{\gamma_a}(\omega_1,\omega_2)$, retaining only the well
    observable features. (d) $Q_{\gamma_a}(\omega_1,\omega_2)$ with a
    black (grey) mask superimposed to let appear only areas where
    $N_c\ge 10^5$ coincidence are obtained in a time
    $T_{\mathrm{exp}}=12$~h (1~h) for~$\gamma_a=1/$ns. Parameters:
    $\gamma_a=0.1g$, $\gamma_\sigma=0.001g$ and
    $P_\sigma=0.01\gamma_a$.}
  \label{fig:1}
\end{figure}

Our main theme in this text is illustrated in Fig.~\ref{fig:1},
starting with the 2PS of the JC (b) under weak incoherent pumping in
the regime of spontaneous emission~\cite{delvalle09a,poshakinskiy14a},
in which case its spectral lineshape is simply the Rabi doublet
(a). The physical meaning of this correlation map is amply discussed
in Ref.~\cite{gonzaleztudela13a}. It is enough for our discussion to
highlight the main phenomenology, namely the set of horizontal and
vertical lines, that correspond to transitions between real states,
and antidiagonal lines $\omega_1+\omega_2=E_{2,\pm}-E_0$ that
correspond to two-photon ``leapfrog'' emission from the second
manifold with levels at energies $E_{2,\pm}$ and the ground state at
energy~$E_0$. The transitions at the Rabi frequency $\pm R_1\approx\pm
g$ are all antibunched (blue on the figure), since they are dominated
by the decay of one polariton from the lower manifold and one
excitation cannot be split into two polaritons, while the lines at
$(\sqrt{2}\pm 1)g$ are mainly bunched (red on the figure), since they
correspond to a cascade from the second manifold. The presence of such
cascade correlations in a regime of low excitations, where the second
manifold has a vanishing probability to be excited, illustrates the
somewhat artificial character of the 2PS. The problem really pertains
to photon-correlations in general rather than to the inclusion of
frequency, since they similarly predicts $g^{(2)}(0)=0$ regardless of
the pumping intensity, that is to say, the system exhibits
antibunching of its overall emission however small is the probability
for two excitations to be present simultaneously, and therefore for
photon blockade to enforce the antibunching~\cite{birnbaum05a}. The
same holds for the harmonic oscillator at vanishing pumping which
still generates bunched statistics $g^{(2)}(0)=2$ regardless of the
probability to reach two excitations in the system. The paradox arises
from the fact that in the limit where the probability of two-photon
effects vanishes, so does the possibility to perform a measurement,
since there is no signal. Instead, if one considers the Mandel
correlations, Eq.~(\ref{Eqmandelfreq}), that are shown in
Fig.~\ref{fig:1}(c), one sees how the result makes more physical
sense: most of the nonlinear features have disappeared or are
considerably weakened in the regions where there is a strong signal
(the correlations tend to die more slowly than the signal), and the
remaining features are concentrated on antibunching between the peaks,
as well as a trace of the bunching cascades. The leapfrog correlations
are extremely strong, which is a general result in all systems, while
the antibunching background that dominates the 2PS profile has now
disappeared. It is also worth noting how the autocorrelation of each
peak (along the diagonal) has the butterfly shape due to
indistinguishability bunching enforced by
filtering~\cite{gonzaleztudela13a}, while cross-correlation between
the two peaks feature a structureless, and therefore neater,
antibunching. This could be of interest for single-photon
emitters~\cite{deutsch12a}.

Although the Mandel correlation spectrum, Fig.~\ref{fig:1}(c), appears
more physical than the underlying 2PS, Fig.~\ref{fig:1}(b), the latter
still presents us with a more fundamental physical picture. Indeed, we
have merely tamed down the features, not removed them, and it is
useful to keep track of correlations even though they are out of reach
of an actual experiment. In any case, the 2PS could still be measured
ideally and should better be regarded as a theoretical limiting
case. The 2PS indeed converges to a unique result in the limit of
vanishing pumping, thereby defining an unambiguous correlation map,
while its Mandel counterpart tends to zero and the relative importance
of bunching versus antibunching areas in Fig.~\ref{fig:1}(c) depend on
one's choice of the pumping rate. Finally, it is worth mentioning that
by the time of writing this text, the 2PS of resonance fluorescence
has already been measured in its entirety~\cite{arXiv_peiris15a} even
for a large splitting of the satellite peaks with spectral windows of
little emission. It seems therefore reasonable that with the
ever-increasing technological progress, all fundamental quantum
optical emitters, even those with much smaller emission rates, will be
likewise characterized.

\section{Valleys of accessible correlations}
\label{sec:vieago15194608CEST2014}
 
While $Q_\Gamma(\omega_1,\omega_2)$ provides a physically sound
picture of which regions of the 2PS are the most
favorable for observation, it also suffers from its own shortcomings.
The arbitrary scale of~$Q$ makes it difficult to attach to it a
quantitative figure of merit.  In this Section, we further delineate
the valleys of accessible correlations based on a down-to-earth
estimate of the numbers of coincidences that can be extracted from the
emission.

Assuming no correlations, the possibility of detecting at least one
coincidence in a time window~$\Delta t$ at frequencies $\omega_i$,
within the frequency windows $\Gamma_i$, is given by:
\begin{equation}
  p=(1-e^{-n_1})(1-e^{-n_2})\,,
\end{equation}
where $n_i=S_{\Gamma_i}(\omega_i) \gamma \Delta t$ represents the
number of filtered photons from a source that emits at a rate
$\gamma$.  For simplicity we always consider symmetric filters in this
Section, $\Gamma_1=\Gamma_2$. Using that definition and assuming no
further inefficiencies in detection, we can estimate the experimental
time required to obtain a given number of random coincidences, $N_c$,
as follows:
\begin{equation}
T_{\mathrm{exp}}= N_c \Delta t /p \,.
\end{equation}
With this, we plot the regions that would be resolved with increasing
experimental time, $T_{\mathrm{exp}}$. For example, assuming
$\gamma_a=1$ ns$^{-1}$ as a quantum dot figure of
merit~\cite{hennessy07a}, we show in Fig.~\ref{fig:1}(d) the
frequency-resolved Mandel parameter with a mask over the regions for
which the number of coincidences is $N_c<10^5$ for a detection time of
$T_{\mathrm{exp}}=1$ (gray) and $12$ hours (black) for $\Delta
t=1/\Gamma$. This shows how the regions with a sizable number of
coincidences reduce to those involving at least one peak, as
expected. Therefore, a first experimental confirmation of these
results could be to keep one branch of the setup on one peak, and
correlate its input with that of the other branch sweeping the entire
spectrum. This should display transitions from antibunching, no
correlation, strong bunching, no correlation again and a weaker
antibunching in the autocorrelation, due to indistinguishability
bunching. For this set of parameters, the experiment would need to run
stably for a longer time in order to collect the same amount of signal
to observe also the leapfrog processes without intersecting with the
peaks. There is a non-trivial interplay of the system parameters that
helps/hinders the observation of correlations which we will be
explored in Section \ref{sec:JC}.

\begin{figure}[!tb]
  \includegraphics[width=0.99\linewidth]{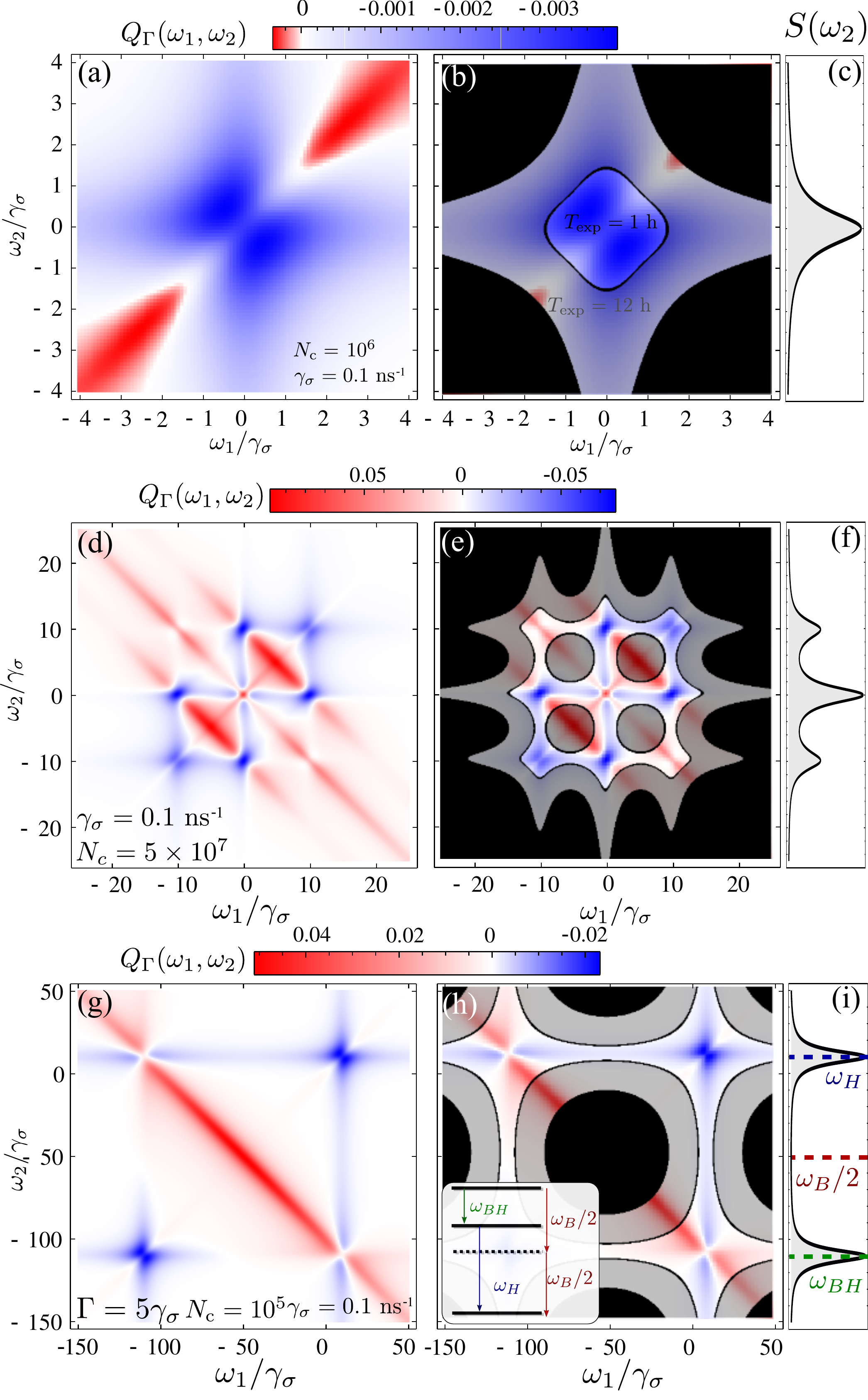}
  \caption{(Colour online) Panel (a-b) [(c)]: Mandel two-photon
    spectra [one-photon spectrum] for an incoherently pumped two-level
    system with $P_\sigma=0.01\gamma_\sigma$, $\gamma_\sigma=0.1$
    ns$^{-1}$ and $\Gamma=\gamma_\sigma$. In Panel (b) we introduce a
    black (grey) mask for the points where $N_c< 10^6$ in a
    time $T_{\mathrm{exp}}=12$ h (1 h). Panel (d-e) [(f)]: Mandel
    two-photon spectra [one-photon spectrum] for a coherently driven
    two-level system with $\Omega_\sigma=5\gamma_\sigma$,
    $\gamma_\sigma=0.1$ ns$^{-1}$ and $\Gamma=\gamma_\sigma$. In Panel
    (e) we introduce a black (grey) mask for the points where $N_c<
    5\times 10^7$ in a time $T_{\mathrm{exp}}=12$ h (1 h). Panel (g-h)
    [(i)]: Mandel two-photon spectra [one-photon spectrum] for an
    incoherently pumped biexciton system with
    $P_\sigma=\gamma_\sigma$, $\chi=100\gamma_{\sigma}$,
    $\gamma_\sigma=0.1$ ns$^{-1}$ and $\Gamma=5\gamma_\sigma$. In
    Panel (h) we introduce a black (grey) mask for the points where
    $N_c< 5\times 10^5$ in a time $T_{\mathrm{exp}}=12$ h (1 h).}
  \label{fig2}
\end{figure}

\begin{figure*}[!htb]
 \includegraphics[width=0.95\linewidth]{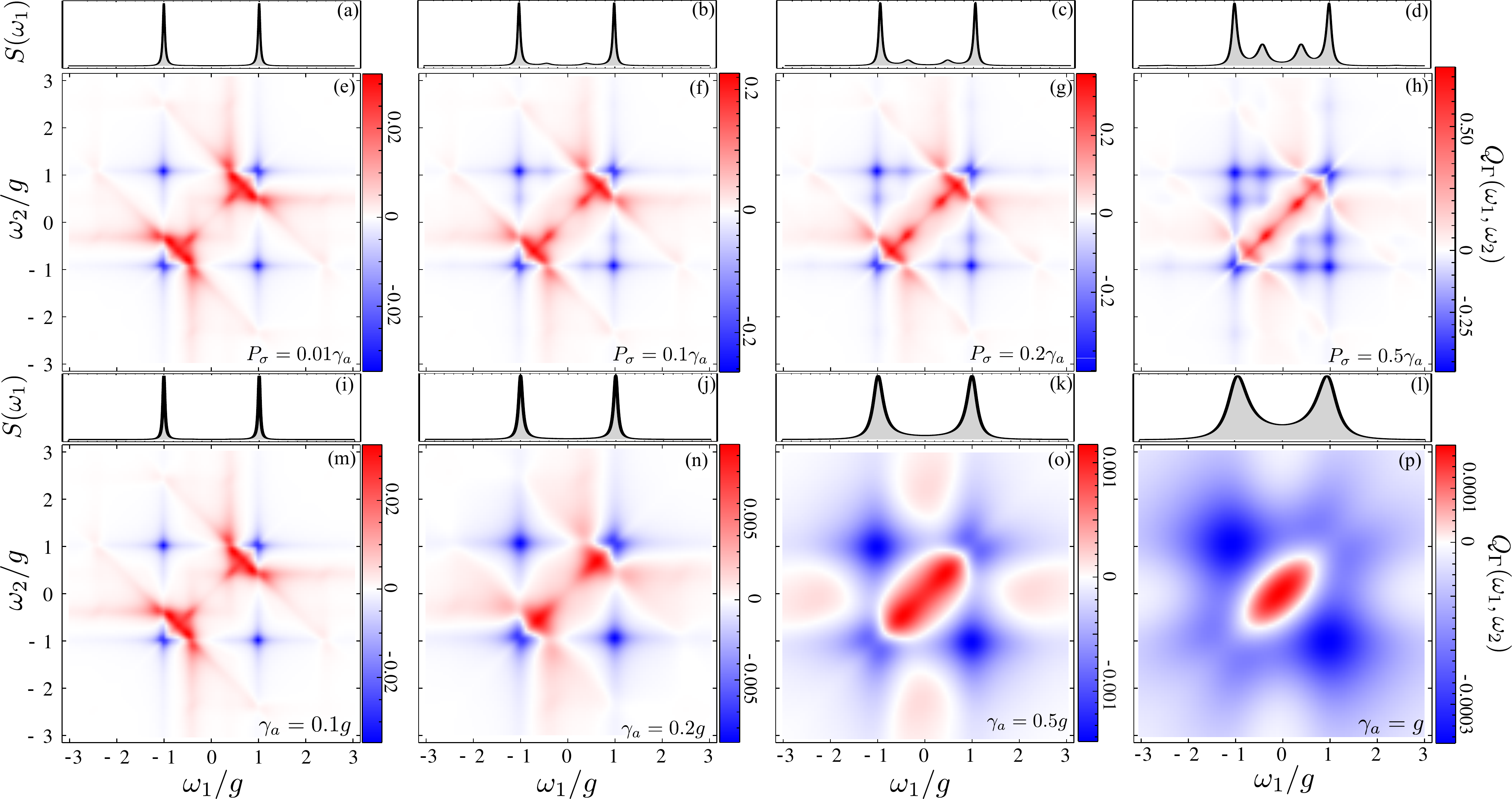}
 \caption{(Colour online) Panels (a-d): One-photon spectra for a JC
   system with $\gamma_a=0.1g$, $\gamma_\sigma=0.001g$ and increasing
   pumping $P_\sigma$ as depicted in the legend. Panels (e-f): Mandel
   two-photon spectra for a JC system with the same parameters as in
   (a-d). Panels (i-l): One-photon spectra for a JC system with
   $P_\sigma=0.01g$, $\gamma_\sigma=0.001g$ and increasing cavity
   decay rate, $\gamma_a$ as depicted in the legend. Panels (m-p):
   Mandel two-photon spectra for a JC system with the same parameters
   as in (i-l).}
  \label{fig3}
\end{figure*}

Before moving on to the optimization, we review other examples of
nonlinear systems explored in Ref.~\cite{gonzaleztudela13a} in the
light of the frequency resolved Mandel parameter and the estimated
time to resolve it. We start by the most basic system that displays a
non-trivial map of correlations, namely, the incoherently pumped
two-level system which we recover by setting $g=0$ in the JC model.
The one-photon spectrum of this system is a single Lorentzian peak
with broadening~$\Gamma_\sigma=\gamma_\sigma+P_\sigma$ as shown in
Fig.~\ref{fig2}(c). Its two-photon Mandel spectrum shows a
\emph{butterfly} shape of anticorrelation, typical of two-level
systems~\cite{gonzaleztudela13a}. By choosing $\Gamma=\gamma_\sigma$
and $\gamma_\sigma=0.1$ ns$^{-1}$, the analysis of the measuring time
shows that a small region of frequency antibunching with $N_c> 10^6$ would be observed within one hour, whereas most of the butterfly
would be observed within $12$ hours for the same threshold of
counts. While it is much more binding to observe, if filtering far in
the tail of a two-level system, one should indeed observe bunching,
against naive expectations.

Next, we consider a resonant coherent driving of the two-level system,
described by Hamiltonian $H_{d}=\Omega_\sigma(\sigma
+\sigma^\dagger)$. In the weak driving regime, the system has been
recently exploited to design ultra-narrow single-photon sources
\cite{nguyen11a,matthiesen12a,he13a,makhonin14a,arXiv_makhonin14a}. In the
strong-driving regime, which is the one that we focus on in this
paper, the spectrum is the well known \emph{Mollow
  triplet}~\cite{mollow69a}, as shown in
Fig.~\ref{fig2}(f). Frequency-resolved correlations for this system
have been theoretically investigated in the
past~\cite{wodkiewicz80a,arnoldus83a,arnoldus84a,nienhuis93a,shatokhin01a}
and even measured~\cite{aspect80a,ulhaq12a,weiler13a} before the
concept of the 2PS was put forward. But in both the theoretical and
experimental contexts, this was at the particular frequencies of the
three peaks. Notwithstanding, interesting correlations arise mainly
outside the peaks~\cite{gonzaleztudela13a,sanchezmunoz14b}, at the
cost of a weaker signal. This has been confirmed
experimentally~\cite{arXiv_peiris15a} with the full reconstruction of
the Mollow 2PS. Resonance fluorescence is indeed a system ideally
suited to pioneer a comprehensive analysis of frequency photon
correlations, since it is obtained in the strong-driving regime of an
extremely quantum emitter, which implies a large emission of strongly
correlated photons.  Figure~\ref{fig2}(e) shows that with
$\gamma_\sigma=0.1$ ns$^{-1}$ and $\Gamma=\gamma_\sigma$, only
within~$1$ hour of experimental time, the regions with $N_c> 5 \times
10^7$ unveils all the horizontal and vertical grid of correlations and
great part of the leapfrogs. It only takes $12$ hours to reveal the
complete two-photon Mandel spectrum.

Finally, we consider a biexciton level scheme as described in
Refs.~\cite{delvalle10a,delvalle11d,delvalle13a} which is relevant in
semiconductor quantum optics as it describes the typical level
structure of quantum dots beyond the simplest two-level system picture
\cite{chen02a,akimov06a,ota11a}. Focusing on a single polarization, it
consist of a three-level scheme as depicted in the inset of
Fig.~\ref{fig2}(h), with a ground, an excitonic (at energy $\omega_H$)
and a biexcitonic state whose energy ($\omega_B$) differs from the sum
of its excitonic constituents by $\chi$ due to Coulomb
interaction. The one-photon spectrum is then composed of two peaks (at
energies $\omega_H$ and $\omega_{BH}=\omega_B-\omega_H$) which give
rise to an interesting and rich landscape of two-photon
correlations. The most prominent feature is the antidiagonal
corresponding to the leapfrog between ground and biexciton state,
$\omega_1+\omega_2=\omega_B=-\chi$ (assuming $\omega_H=0$ as the
reference energy), with potential for applications in the generation
of entangled photon pairs by frequency
filtering~\cite{delvalle13a}. With~$\gamma_\sigma=0.1$ ns$^{-1}$ and
setting the threshold at $N_c=10^5$ random coincidences, within one
hour, the anticorrelation area and bunching of the one-photon
transition peaks would be observable, whereas in $12$ hours most of
its leapfrog structure would be revealed as well, especially by
filtering on the sides of the one-photon peaks. Due to both its
fundamental importance and practical applications, we will return to
the problem of optimizing the observation of the leapfrog processes by
changing both the intrinsic parameters as well as the filtering ones
in Section~\ref{sec:biexciton}.

\section{Optimization of correlations in the Jaynes-Cummings model}
\label{sec:JC}

We now illustrate how to optimize photon correlations thanks to
frequency filtering, in the particular case of the JC model. We do a
qualitative analysis to avoid focusing the discussion on a particular
set of experimental figures of merit.  We consider the system
parameters are variables for the optimization and postpone to next
Section (and to other systems) the optimization through extrinsic
parameters, e.g., the filters and detection time.

One parameter that can be easily modified is the incoherent pump rate,
$P_\sigma$. The example of the previous Section was chosen to be well
into the linear regime, i.e., with a very small pumping rate, namely
$P_\sigma=0.01\gamma_a$. Increasing pumping has two interesting
consequences for the observation of correlations:
\begin{enumerate}
\addtolength{\itemsep}{-0.5\baselineskip}
\item signal increases,
\item the system enters the nonlinear regime.
\end{enumerate}

In Fig. \ref{fig3}, upper row, the effect of increasing pumping is
shown for both the spectral shape (a--d) and the Mandel parameter
resolved in frequency (e--h). The spectra let appear inner peaks
between the Rabi doublet, corresponding to transitions from the
higher manifolds. The corresponding Mandel 2PS also develops new
features at the same time as the overall intensity of the correlations
increases, from $\sim 0.02$ with $P_\sigma=0.01\gamma_a$ to $\sim 0.5$
with $P_\sigma=0.5\gamma_a$ (note the change of the color scales). In
particular, the higher manifolds become visible in antibunching only
as they get populated, while they manifest themselves in bunching more
clearly at low pumping.  The Jaynes--Cummings fork provides a
well-structured set of correlations between the peaks: the inner peaks
emit bunched photons but are otherwise antibunched with each other,
or with the remotest Rabi peak, and are uncorrelated with the
other---closer---Rabi peak. When various manifolds are well populated,
correlations are dominated by real-state transitions and virtual
processes shy away in comparison.

\begin{figure}[!tb]
 \includegraphics[width=0.8\linewidth]{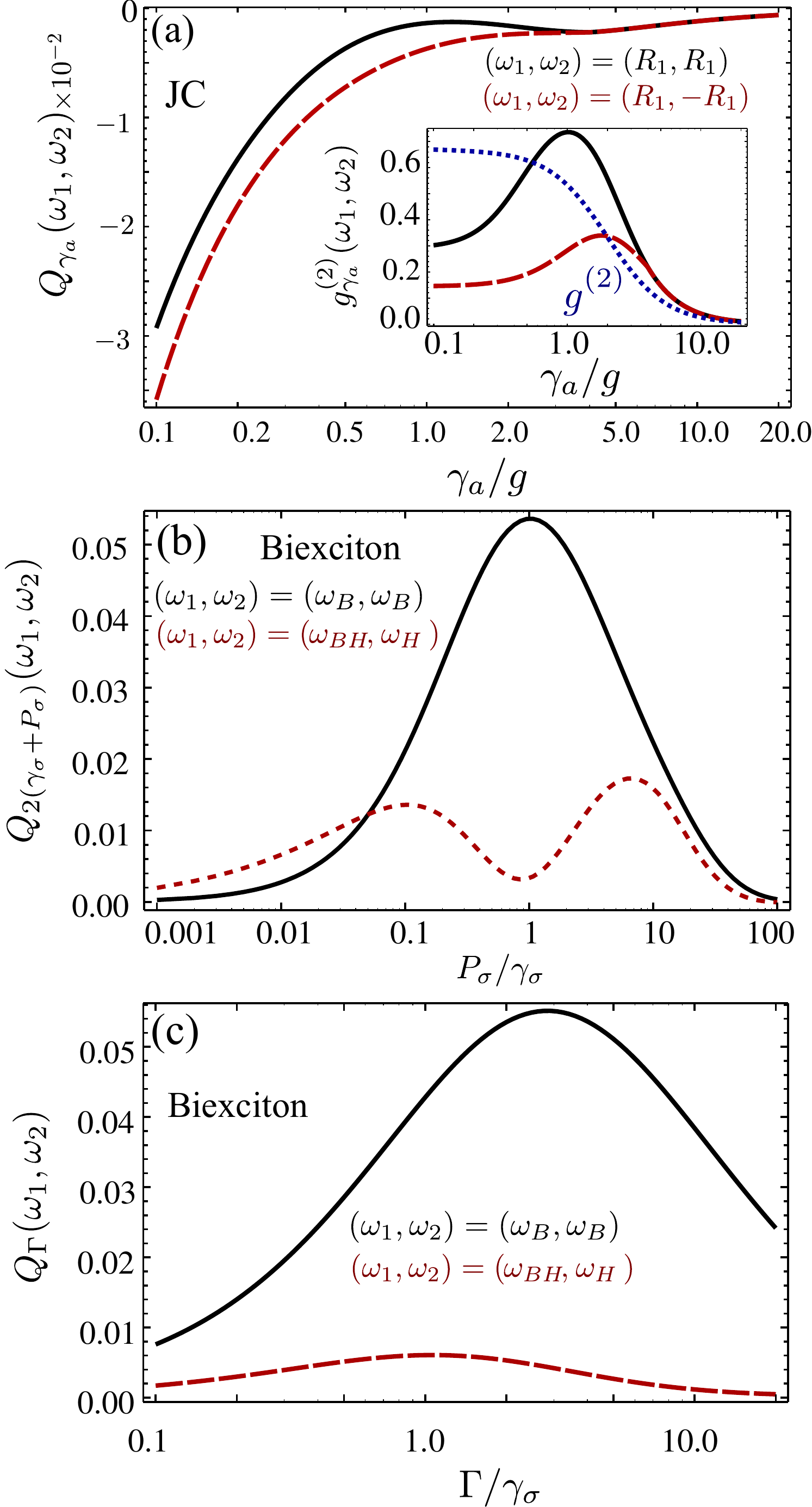}
 \caption{(Colour online) (a) $Q_{\gamma_a}(\omega_1,\omega_2)$ as a
   function of $\gamma_a$ for a JC system with
   $P_\sigma=0.01\gamma_a$, $\gamma_\sigma=0.001g$ for pair of
   frequencies $(R_1,R_1)$ (solid black), $(R_1,-R_1)$ (dashed
   red). Inset: $g_{\gamma_a}^{(2)}(\omega_1,\omega_2)$ for the same
   parameters, together with standard photon correlation $g^{(2)}(0)$
   (dotted blue). (b) $Q_{\Gamma}(\omega_1,\omega_2)$ as a function of
   $P_\sigma$ for a biexciton cascade with the same parameters as in
   Fig. \ref{fig2} and $\Gamma=2(\gamma_\sigma+P_\sigma)$, for pair a
   of frequencies as indicated in inset. (c)
   $Q_{\Gamma}(\omega_1,\omega_2)$ as a function of $\Gamma$, with the
   same parameter of Fig. \ref{fig2}.}
 \label{fig4}
\end{figure}

Another parameter that is less easily tuned but that determines the
strong-coupling property is the cavity decay rate. In the bottom row
of Fig.~\ref{fig3}, the effect of increasing $\gamma_a$ is shown,
always keeping the system in the strong-coupling regime ($\gamma_a<4
g$). This results in the structure smoothing out as well as the
intensity of correlations dying (note, again, the color scales). The
absolute scale of the Mandel correlations indeed decreases by one
order of magnitude, but partly because the cavity gets less populated,
and increasing pumping could compensate for that. For $\gamma_a=0.5g$
the leapfrog antidiagonal have disappeared and for $\gamma_a=g$, only
the anticorrelation between the Rabi peaks have survived, surrounding
a region of indistinguishability bunching.  To track more
quantitatively how the frequency-resolved correlations evolve with the
cavity decay rate, we show in Fig.~\ref{fig4}(a) the value of
$Q_{\gamma_a}$ for pairs of frequencies corresponding to filtering the
Rabi peaks. To show that there is some difference due to
indistinguishability bunching in the case of autocorrelation, we
present both the filtering for the same Rabi peak (in solid black) and
for the two Rabi peaks (dashed red). For $\gamma_a\ge 4 g$, when the
system reaches the weak-coupling regime, frequency-filtered
correlations collapse into a single curve since there are no longer
polariton modes in the system. It is instructive in this case to plot
$g^{(2)}_{\gamma_a}(R_1,\pm R_1)$ together with the standard $g^{(2)}$
(dotted blue) as shown in the inset of Fig.~\ref{fig4}(a). The
difference in this case between filtering the same peak or
cross-correlating them becomes significant.  As already observed,
frequency filtering the Rabi peaks improves antibunching as compared
to non-filtered correlations, as we are discarding the frequency
regions with bunched photons. This is another instance of how
frequency filtering can be used to harness correlations. Note also
that worsening the cavity quality factor betters the overall
antibunching while it spoils the peaks antibunching.
 
\section{Optimization of correlations in a biexciton cascade}
\label{sec:biexciton}

We now study photon correlations in a biexciton cascade. To link with
the previous discussion on the JC, we show in Fig.~\ref{fig4}(b) the
dependence on the pumping rate for two configurations, letting here the frequency window grow with the pumping as $\Gamma=2(\gamma_\sigma+P_\sigma)$. The two
configurations are filtering the leapfrog transition at the biexciton
frequency (in solid black) and the biexciton-exciton cascade (in
dashed red). Both transitions are bunched, $Q_\Gamma>0$, due to their
two-photon cascade character. However, the one that corresponds to the
leapfrog process exhibits a clear optimal pumping intensity, at
$P_\sigma\sim 1.5\gamma_\sigma$, whereas the one-photon transitions
exhibit two local maxima at around $P_\sigma\sim 0.2\gamma_\sigma$ and
$P_\sigma\sim 8\gamma_\sigma$, that follow the successive growth of
the exciton and biexciton populations. The low pumping regime leads to
$Q_\Gamma\rightarrow 0$ due to the small emission whereas in the high
pumping case this is because one recovers the standard photon
correlations $g^{(2)}(0)\approx 1$.

\subsection{Asymmetric filters}

\begin{figure}[!tb]
  \includegraphics[width=0.95\linewidth]{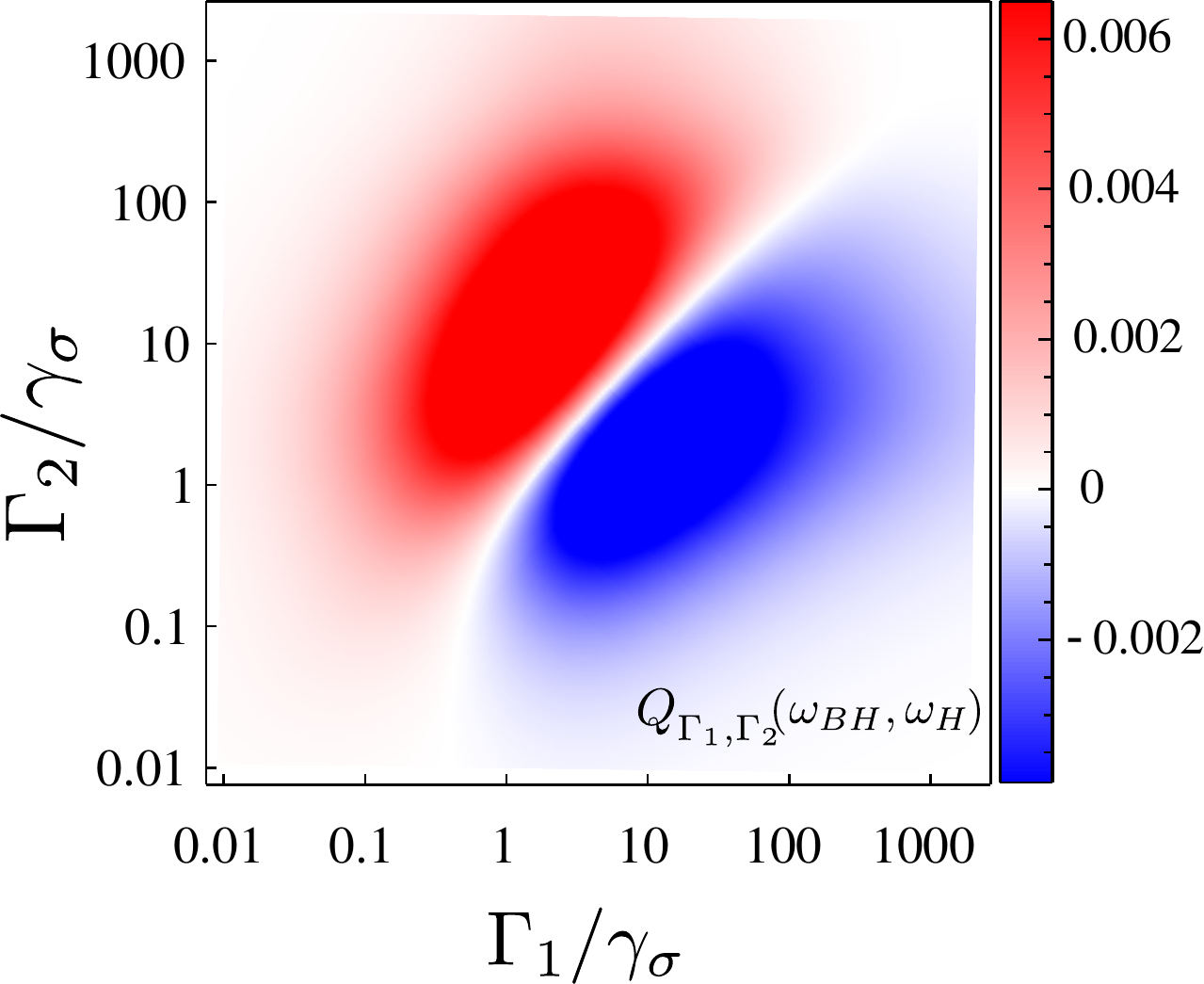}
  \caption{(Colour online) Mandel parameter
    $Q_{\Gamma_1,\Gamma_2}(\omega_{BH},\omega_{H})$ when filtering the
    two peaks of the biexciton-exciton cascade as a function of the
    widths of the spectral windows. The diagonal corresponds to the
    typical case of identical filters. Stronger correlations and of a
    varying nature are however obtained for asymmetric
    filters. Parameters are the same as for Fig.~\ref{fig2}. }
  \label{fig5}
\end{figure}
 
We discuss in more details how the filter linewidth affects the
correlations. In Fig.~\ref{fig4}(c), we show the dependence with the
filters linewidth $\Gamma$ for correlations of the leapfrog cascade
(solid black) and the biexciton-exciton cascade (dashed red). The
leapfrog correlations strongly depend on the filter linewidth, due to
the virtual nature of their transitions, with an optimum value at
around $4\gamma_\sigma$. The biexciton-exciton cascade also displays a
maximum $\Gamma\sim \gamma_\sigma$, but its dependence is much
weaker. In both cases, and in contrast with
$g^{(2)}_{\Gamma}(\omega_1,\omega_2)$ where smaller linewidths
optimize leapfrog correlations, the compromise for signal requires
larger filter linewidths to optimize the intensity of correlations.

In the analysis so far, we have always considered symmetric filters
for the two frequencies: $\Gamma_1=\Gamma_2=\Gamma$.  However, for a
cascaded emission such as
biexciton~$\rightarrow$~exciton~$\rightarrow$~ground state, it is
worth exploring the situation where the filters are asymmetric,
$\Gamma_1\neq \Gamma_2$. This is shown in Fig.~\ref{fig5}, filtering
the biexciton/exciton peaks for a situation where they have the same
broadening and intensity ($P_\sigma=\gamma_\sigma$).  Two areas of
bunching/antibunching oppose each other depending on the relative
value of $\Gamma_1$ vs $\Gamma_2$, separated by a frontier of
no-correlation that roughly correspond to the case of symmetric
filters, showing the interest in lifting this limitation even when
both spectral peaks are equal.  Such a structure is typical of photon
cascades. From the level structure, the natural order of the cascade
makes it indeed likely to detect a photon first of frequency
$\omega_{BH}$ then of $\omega_H$.  Since an $\omega_{BH}$
[$\omega_{H}$] photon, filtered with $\Gamma_1$ [$\Gamma_2$], is the
first [second] photon in the cascade, if $\Gamma_2>\Gamma_1$ the time
spent by the photon in filter $\omega_1$ is larger than the one
$\omega_2$ which is favouring the simultaneous detection of the two photons of the cascade, and
therefore, yields a strong bunching. In the opposite regime, when
$\Gamma_1>\Gamma_2$, the $\omega_{BH}$ photon spends less time in the
filter preventing the simultaneous detection of the two photons of the cascade and therefore
yielding strong antibunching in the Mandel parameter. There is an optimum
value to observe either antibunching/bunching---as a rule of thumb, an
order of magnitude difference---since ultimately the observations of
correlations quenches for very broad/asymmetric filters. This loss of
correlations is due, interestingly, to the overlap of the filtering
windows.

\subsection{Delayed correlations}

\begin{figure}[!tb]
  \includegraphics[width=0.9\linewidth]{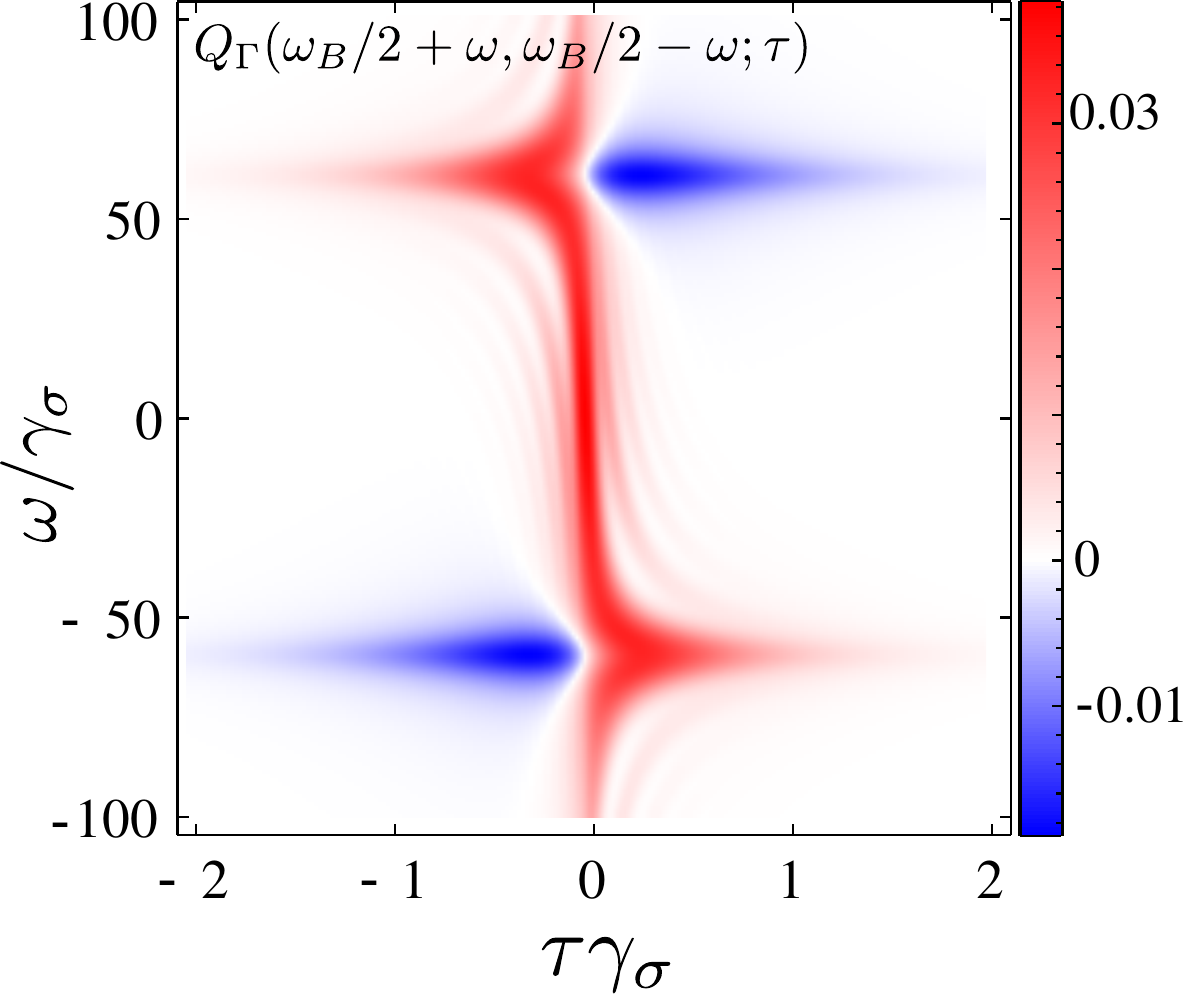}
  \caption{(Colour online) Mandel parameter
    $Q_{\Gamma}(\omega_B/2+\omega,\omega_B/2-\omega;\tau)$ when
    filtering the leapfrog processes (antidiagonal) of the
    biexciton-exciton cascade and as a function of autocorrelation
    time. This shows the contrast between leapfrog (virtual) processes
    where the correlations are symmetric in~$\tau$ and with a fast
    decay, and real processes that, when intercepted by the filters,
    lead to characteristic antibunching-bunching transitions with a
    slow decay.  Parameters are the same as in Fig. \ref{fig2} except
    for $\Gamma=10\gamma_\sigma$. }
  \label{fig6}
\end{figure}
 
We have also restricted our attention to $\tau=0$, i.e.,
coincidences. However, particularly for cascaded emission, it is to be
expected that correlations are maximum at nonzero
delay~\cite{delvalle12a}.  To condense the bulk of the information
into a single figure (Fig.~\ref{fig6}), we consider the joint
frequency- and time-resolved Mandel 2PS along the leapfrog
antidiagonal of Fig.~\ref{fig2} (e-f),
$Q_{\Gamma}(\omega_B/2+\omega,\omega_B/2-\omega;\tau))$ for
$\Gamma=10\gamma_\sigma$.  In this line lies the information of both
the leapfrogs and the one-photon cascade.  Leapfrog emission is
maximum at $\omega=0$~\cite{delvalle13a} and is symmetric in $\tau$,
which is typical of second-order processes where the photons, being
virtual, have no time order. Due to this symmetry, the optimal delay
to observe correlations in this case is $\tau=0$, and they decay with
the filter timescale~$1/\Gamma$. Contrarily, the biexciton-exciton
photon cascade, appearing at $\omega=\{\omega_{BH},\omega_{H}\}$, is
strongly asymmetric, as clearly shown in Fig. \ref{fig6}.  It shows a
bunching/antibunching behaviour as expected for cascades of real
transitions, where there is a definite temporal order in the
emission. In this case, the optimal value to observe strong
bunching/antibunching is $\tau\sim 1/\Gamma$ and the correlations
ultimately decay in the intrinsic timescale of the system given by
$1/\gamma_\sigma$. The different timescales where correlations survive
between leapfrog and normal cascaded emissions is another consequence
of their different physical origin.

\subsection{Combining the parameters}

\begin{figure}[!tb]
  \includegraphics[width=0.9\linewidth]{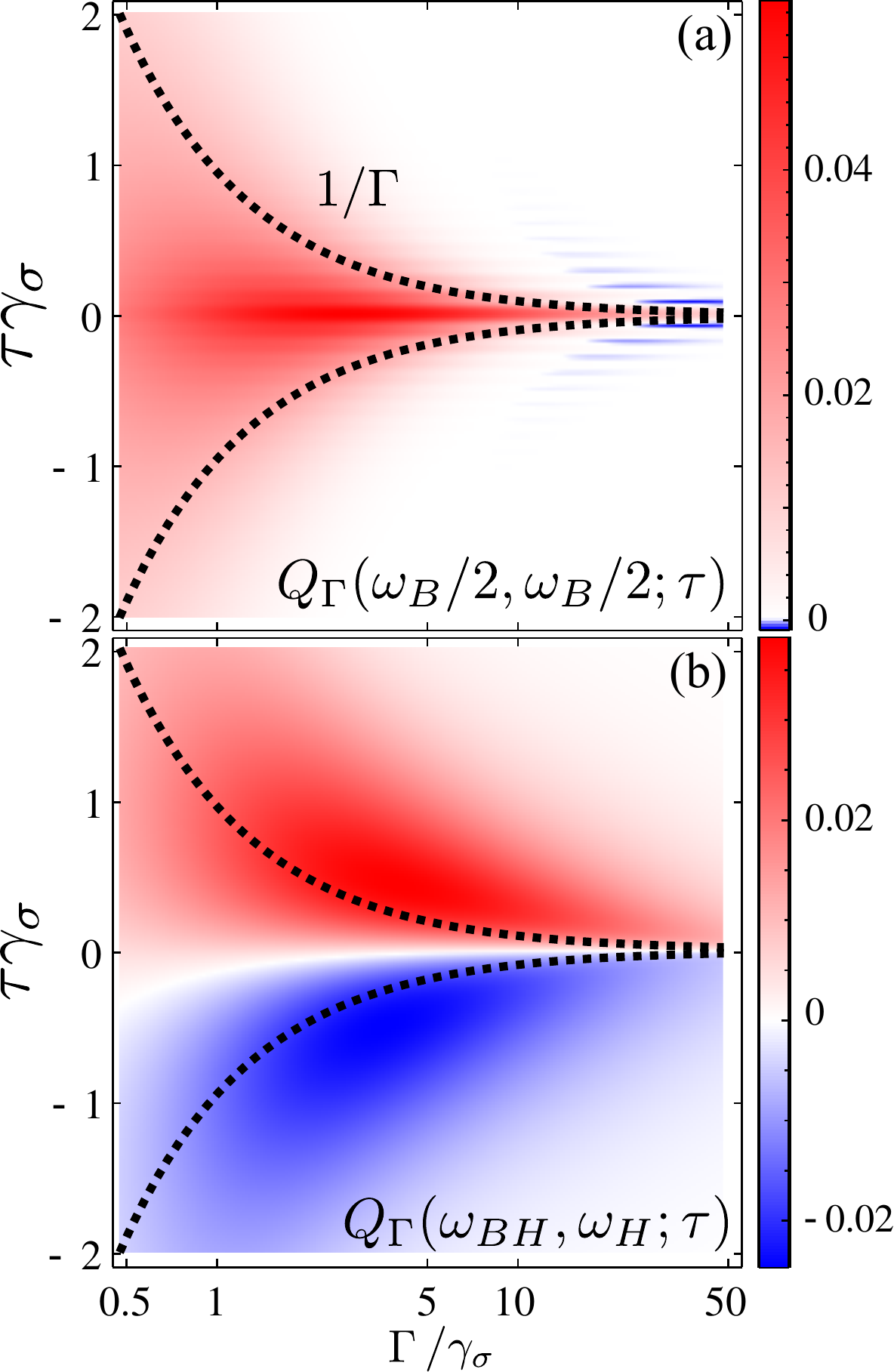}
  \caption{(Colour online) (a) Mandel parameter
    $Q_{\Gamma}(\omega_B/2,\omega_B/2;\tau)$ when filtering the
    leapfrog processes of the biexciton-exciton cascade and (b)
    $Q_{\Gamma}(\omega_{BH},\omega_{H};\tau)$ when filtering the two
    peaks, as a function of correlation time~$\tau$ and filters
    linewidth~$\Gamma$.  This highlights again the difference between
    real and virtual processes. An optimum value of the filters
    linewidth can be found for the case of real state
    transitions. Parameters are the same as for Fig.~\ref{fig2}.}
  \label{fig7}
\end{figure}

Finally, after having explored separately both the dependence on the
filter linewidth, $\Gamma$, and the temporal delay of the photons,
$\tau$, one can naturally think of combining them to optimize
correlations cumulatively. In Fig.~\ref{fig7}, we show the joint
$\tau$ and $\Gamma$ dependence of the frequency-resolved Mandel
parameter for the two more relevant cases: the leapfrog two-photon
cascade in~(a) and the consecutive one-photon transitions in~(b).  As
previously discussed, the temporal shape of the leapfrog cascade is a
symmetric decay of correlations within timescale $1/\Gamma$, as shown
in the figure.  As $\Gamma$ increases, so does the decay rate of the
correlations (the correlation time and the filters linewidth are
roughly in inverse proportion, as shown by the dotted lines which are
$1/(\Gamma/\gamma_\sigma)$). The $\tau=0$ correlation also strongly
decreases and is eventually surrounded by antibunching oscillations in
$\tau$, that correspond to the fast off-resonance one-photon
transitions.  For the set of parameters chosen here, the optimal
correlations are to be found at $\tau=0$ and for $\Gamma \sim 3
\gamma_\sigma$. Also as discussed previously, in the consecutive
one-photon cascade in Fig.~\ref{fig7}(b), the temporal shape exhibits
a typical asymmetric bunching/antibunching shape. The pattern is
fairly robust but can indeed be magnified by the appropriate choice of
filters (we consider here symmetric filters for simplicity). The
temporal decay occurs this time approximately within a time scale of
$1/\gamma_\sigma$, while the maximum value for the correlations, both
bunching and antibunching, is obtained at $\tau \sim 1/\Gamma$. In
this case, the maximum is found at $\Gamma\sim 3 \gamma_\sigma$ and
$\tau\sim1/\Gamma$. Note that correlations are optimised for filters
with a width equal to the spectral peaks ($3\gamma_\sigma$ for our
choice of parameters).

\section{Summary and Conclusions}

Summing up, filtering the photons emitted by a quantum source has a
dramatic impact on their correlations. Strong correlations are often
found in regions of the spectrum where there is a weak emission,
making their experimental detection particularly difficult, since this
implies coincidences of rare events. We have introduced a
``frequency-resolved Mandel parameter'' as well as a quantitative
estimate of the time required to accumulate a given number of
coincidences, to address this problematic for several paradigmatic
non-linear quantum systems. We have shown the considerable flexibility
opened by frequency-filtering, either in energy or in time, with
possibilities to enhance correlations by varying filters linewidths
(possibly asymmetrically), temporal windows of detections and system
parameters (such as pumping). Depending on whether the correlations
originate from real states transitions or involve virtual processes,
different strategies should be adapted, corresponding to their
intrinsically different character. With the recent experimental
demonstration of the underlying physics~\cite{arXiv_peiris15a}, the
field of two-photon spectroscopy is now ripe to power applications and
optimize resources based on photon correlations.

\begin{acknowledgements}
  This work has been funded by the ERC Grant POLAFLOW. EdV
  acknowledges support from the IEF project SQUIRREL (623708) and AGT
  from the Alexander Von Humboldt Foundation.
\end{acknowledgements}

\bibliography{Sci,books,arXiv}

\end{document}